\pdfoutput=1
\documentclass{article}

\usepackage[english]{babel}
\usepackage[letterpaper,top=2cm,bottom=2cm,left=3cm,right=3cm,
marginparwidth=1.75cm]{geometry}
\usepackage[colorlinks=true, allcolors=blue]{hyperref}

\usepackage[utf8]{inputenc}
\usepackage{graphicx}
\usepackage{amsmath}
\usepackage{amssymb}
\usepackage{mathtools}
\usepackage{color}
\makeatletter \@addtoreset{equation}{section} \makeatother

\newcommand{\cf}{cf.\ }
\newcommand{\eg}{e.g.\ }
\newcommand{\ie}{i.e.\ }

\newcommand{\wrt}{{with respect to }}

\newcommand{\rhs}{{right--hand side }}
\newcommand{\noi}{\vspace{12pt}\noindent}
\newcommand{\e}[1]{{(\ref{#1})}}
\newcommand{\eq}[1]{{eq.\ (\ref{#1})}}
\newcommand{\es}[2]{{(\ref{#1}) and (\ref{#2})}}
\newcommand{\eqs}[2]{{eqs.\ (\ref{#1}) and (\ref{#2})}}

\newcommand{\equi}[1]{\stackrel{{#1}}{=}}
\newcommand{\beq}{\begin{equation}}
\newcommand{\eeq}{\end{equation}}
\newcommand{\bea}{\begin{eqnarray}}
\newcommand{\eea}{\end{eqnarray}}

\newcommand{\for}{{\rm for}}

\renewcommand{\~}{ \ }
\renewcommand{\=}{ \ = \ }
\renewcommand{\-}{ \!-\! }
\renewcommand{\Bar}{\overline}

\newcommand{\Wf}{f}
\newcommand{\Wi}{i}

\begin{document}
\thispagestyle{empty}
\title{\Large{\bf A Note on Instantons in a 1D\\
Same-Level Asymmetric Double Well}}
\author{{\sc Klaus~Bering}$^{a}$ \\~\\
Institute for Theoretical Physics \& Astrophysics\\
Masaryk University\\Kotl\'a\v{r}sk\'a 2\\CZ--611 37 Brno\\Czech Republic}
\maketitle
\vfill
\begin{abstract}
We prove formulas for the multi-instanton corrections to the overlap and
energies of a 1D same-level asymmetric double well using the Euclidean path
integral. Both the odd and even instanton sectors are summed to all orders.
The double well is same-level asymmetric in the sense that the potentials at
neighboring wells have the same bottom level but can have different
Hessians/curvatures/frequencies, which modify Coleman's original formulas.
This for instance implies that the reference model used to calculate the
functional determinant of quantum fluctuations must now interpolate between
simple harmonic oscillators of different frequencies. Examples of symmetric
double and triple wells are worked out.
\end{abstract}
\vfill
\begin{quote}
\hrule width 5.cm \vskip 2.mm \noindent 
$^{a}${\small E--mail:~{\tt bering@physics.muni.cz}} \\
\end{quote}

% determine the unit for the size of diagram
\unitlength = 1pt

\newpage
\tableofcontents

\section{Introduction}

\noi
In this study, we extend the seminal treatment by Coleman of a dilute 1D
instanton gas via the Euclidean path integral 
\cite{coleman85,vzns82,rajaraman87,altlandsimons10,rattazzi11,paranjape22}
to the case where the various potential wells have different characteristic
frequencies \e{freq01} given by their Hessians/simple harmonic oscillator
(SHO) approximations \cite{bmw94,dunne00,rivero02,vybornyi14,jlr20}. 

\noi
The different frequencies lead to extra multiplicative factors in the
instanton formulas, see \eg \eqs{positionenergyoverlap01}{kay01}. 

\noi
The path integral is calculated as a telescoping product where one divides 
and multiplies with the same reference model factor. The relative functional 
determinant is evaluated via the Gelfand-Yaglom (GY) theorem.

\noi
Another new feature is that the different frequencies force us to modify the
reference model --- from a time-independent --- to a step-wise time-dependent
--- reference potential \e{refpot01} that asymptotically mimics the different
characteristic frequencies at the beginning and end of the instanton's life
in order to comply with the settings of the Gelfand-Yaglom (GY) theorem.

\noi
Remarkably, the all-order multi-instanton corrections (of the Euclidean path
integral of the asymmetric double well) deform the 2 lowest energy states in
a manner that preserves the interpretation as a 2-level system, as it should,
\cf \eqs{eueodd}{eueeven}. This is easiest to see via a Laplace transform from
time domain to energy domain,
\cf \eqs{laplacetransfodd01}{laplacetransfeven01}.

\noi
We will still assume that the bottoms of the different wells in the potential 
$V$ are at approximately the same level, so that instantons do not have to 
dump/gain classical energy. Note that in Coleman's original setting the 
same levels were a natural consequence of the $\mathbb{Z}_2$-symmetry of the 
model; in our asymmetric setting this is an additional assumption, which is 
technically unnatural in the sense of 't Hooft. Nevertheless, one could still
envision fine tuned applications of this.

\noi
There exist other methods that we shall not use in this paper, such as,
\eg the WKB connection formulas \cite{berrymount72}, and methods of complex
integration contours \cite{zinnjustin21,marino10}.
Unlike the Bogomolny-Zinn-Justin (BZJ) prescription \cite{bogo80,zinn81},
which is based on an action for an instanton-anti-instanton-pair/bounce,
this paper treats each instanton transition individually.
We also shall not discuss the related topic of decay rates \cite{cc77,affs16}.

\noi
The paper is organized as follows. In Section~\ref{classec}, we discuss the
classical model. The main multi-instanton formulas are presented in
Sections~\ref{qmsec}-\ref{oddevensec}. The functional determinant of a
$1$-instanton is discussed in Sections~\ref{1inst}-\ref{relfuncdetsec}.
Section~\ref{exsec} contains examples of symmetric double and triple wells.

\subsection{Notation} 

\begin{itemize}
\item 
Capital letters denote the full system. Small letters are used for quantum
fluctuations/SHO well. The overbar denotes a classical instanton solution. 
\item 
The subscripts ``$+$" (``$-$") and ``$f$" (``$i$") denote the final
(initial) time, respectively. 
\item 
The reference model is decorated with an ornament ``$(0)$".
\end{itemize}

\section{Classical model}
\label{classec}

\noi
In this Section~\ref{classec} we review the classical 1D model.
The off-shell Euclidean action functional adds (as opposed to subtracts) the
potential term
\beq
S[X]\=\int_{\tau_i}^{\tau_f}\! d\tau \left(\frac{1}{2}\dot{X}^2+V(X)\right).
\label{action01}
\eeq
(We set the mass $m=1$ for simplicity.) The position variable 
\beq
X\=\Bar{X}+x\label{xxx01}
\eeq
is separated into a classical instanton solution $\Bar{X}$ and a quantum
fluctuation $x$. 

\subsection{On-shell action}
\label{actionsec}

\noi
We will assume that the different wells in the potential $V$ are at
approximately the same level (so that instantons do not have to dump/gain
classical energy, \ie the instanton velocity vanishes asymptotically as it
approaches the bottom of a well)\footnote{Note that even before taking
instanton corrections into account, the energy degeneracy is typically
already lifted by the different well's different quantum mechanical
zero-point energies \e{freq01}, \cf next Section~\ref{qmsec}.}.
We will shift the zero-level of the potential so that
the well bottoms have zero potential. Recall that the Euclidean energy
subtracts (as opposed to adds) the potential energy $V$. Euclidean energy
conservation leads to formula \e{energycons01} for the on-shell action
\bea
0&\approx&\Bar{E}\= \frac{1}{2}\dot{\Bar{X}}^2-V(\Bar{X})
\label{energycons00}\cr
&\Downarrow&\cr
0&\leq&S[\Bar{X}]
\equi{\e{action01}}
\int_{\tau_i}^{\tau_f}\! d\tau\left(\frac{1}{2}\dot{\Bar{X}}^2+V(\Bar{X})\right)\cr
&\approx&\Bar{S}\~:=\~\int_{\tau_i}^{\tau_f}\! d\tau \dot{\Bar{X}}^2
\=\int_{X_i}^{X_f}\! d\Bar{X} \~\dot{\Bar{X}}\label{sbar01}\cr
&\approx&\int_{\min(X_i,X_f)}^{\max(X_i,X_f)}\! d\Bar{X} \sqrt{2V(\Bar{X})}.
\label{energycons01}
\eea

\subsection{Asymptotic behavior}

\noi
The Euler-Lagrange (EL) equation implies a zero-mode $\dot{\Bar{X}}$ for the
pertinent differential operator:
\bea
\ddot{\Bar{X}}&\equi{\e{action01}} &V^{\prime}(\Bar{X})
\~\stackrel{\text{SHO}}{\approx}\~
\underbrace{V^{\prime\prime}(X_{\pm})}_{=:\~\omega_{\pm}\~>\~0}(\Bar{X}-X_{\pm})
\label{el00} \cr
&\Downarrow&\cr
\left\{-\partial^2_{\tau} +V^{\prime\prime}(\Bar{X}) \right\}
\dot{\Bar{X}}&=&0.\label{el01}
\eea
In \eq{el01}, the subscript ``$\pm$'' refers to the asymptotic final/initial
(``$f/i$'') time. There is a 1-parameter family of 1-instantons
$\Bar{X}=\Bar{X}_{\tau_1}$ parametrized by the instanton transition
time/collective coordinate $\tau_1$. Let
\beq\Delta\tau_1\~:=\~\tau-\tau_1\label{deltatau101}
\eeq
denote the time measured relative to the instanton collective coordinate
$\tau_1$.

\noi
The asymptotic instanton solutions are exponentially growing/decaying
\e{asymp01} for early/late times $\omega_{\pm}|\Delta\tau_1|\gg 1$, respectively:
\bea
\left|\frac{d\Bar{X}}{d\tau}\right|\=\left|\dot{\Bar{X}}\right|
&\stackrel{\e{energycons01}}{\approx}&\sqrt{2V(\Bar{X})}
\~\stackrel{\text{SHO}}{\approx}\~
\sqrt{V^{\prime\prime}(X_{\pm})(\Bar{X}-X_{\pm})^2}
\~\equi{\e{el00}}\~ \omega_{\pm} |\Bar{X}-X_{\pm}| \cr
&\Downarrow&\e{el00}\cr
d\ln|\Bar{X}-X_{\pm}|\=\frac{d\Bar{X}}{\Bar{X}-X_{\pm}}
&\approx& \mp\omega_{\pm}d\tau\cr
&\Downarrow&\cr
\exists C_{\pm}\~\geq\~0:\qquad |\Bar{X}(\tau)-X_{\pm}|
&\approx& C_{\pm}e^{-\omega_{\pm}|\Delta\tau_1|}.
\label{asymp01}
\eea
In \eq{asymp01} the integration constants $C_{\pm}$ are asymptotic
amplitudes for the instanton.

\section{Quantum mechanical path integral}
\label{qmsec}

\noi
Next, we consider the quantum mechanical (QM) model. Let $U(\tau_f,\tau_i)$
denote the time evolution operator. We assume that each
potential minimum is sufficiently deep such that its simple harmonic oscillator
(SHO) approximation suffices in a neighborhood. (For explicit examples of
this assumption, see \eqs{2pert01}{3pert01}.) In other words, it is
assumed that in periods where the system is not undergoing brief instanton
transitions, it is adequately described as being in a SHO ground state.
In particular, the quantum mechanical path integral/overlap between an
initial (``$i$") and a final (``$f$") well minima
\bea
\langle X_f|U(\tau_f,\tau_i)|X_i\rangle&=&\int_{X(\tau_i)=X_i}^{X(\tau_f)=X_f}\!
[dX]\~e^{-\frac{1}{\hbar}S[X]}\cr
&=&{}_{\Wf}\langle x_f\!=\!0|U(\tau_f,\tau_i)|x_i\!=\!0\rangle_{\Wi}\cr
&=&\sum_{e_f,e_i}{}_{\Wf}\langle x_f\!=\!0|e_f\rangle_{\Wf}
\~{}_{\Wf}\langle e_f|U(\tau_f,\tau_i)|e_i\rangle_{\Wi}
\~{}_{\Wi}\langle e_i|x_i\!=\!0\rangle_{\Wi}\cr
&\stackrel{\e{gaussianwavefct01}}{\approx}&
\left(\frac{\omega_f}{\pi\hbar}\right)^{1/4}
\~{}_{\Wf}\langle e_f|U(\tau_f,\tau_i)|e_i\rangle_{\Wi}
\~\left(\frac{\omega_i}{\pi\hbar}\right)^{1/4},
\label{positionenergyoverlap01}
\eea
is dominated by the quantum mechanical SHO ground state energies 
\beq
e_i\=\frac{\hbar\omega_i}{2}
\qquad\text{and}\qquad
e_f\=\frac{\hbar\omega_f}{2}.\label{freq01}
\eeq
which are generically {\em not} assumed to be the same, in contrast to the
classical energy levels, \cf Subsection~\ref{actionsec}. 
In \eq{positionenergyoverlap01} the overlap
\bea
{}_{\Wi}\langle x_i|e_i\rangle_{\Wi}
&=&\left(\frac{\omega_i}{\pi\hbar}\right)^{1/4}
e^{-\frac{1}{2\hbar}\omega_ix_i^2},\cr
{}_{\Wf}\langle x_f|e_f\rangle_{\Wf}
&=&\left(\frac{\omega_f}{\pi\hbar}\right)^{1/4}
e^{-\frac{1}{2\hbar}\omega_fx_f^2},
\label{gaussianwavefct01}
\eea
between position and energy in the ``$i/f$'' well is modelled by a
Gaussian SHO ground state wavefunction.

\subsection{1-instanton}
\label{1instsec}

\noi
The stationary phase/WKB approximation yields that a 1-instanton
contribution is of the form
\bea
{}_{\Wf}\langle x_f\!=\!0|U_{\tau_1}(\tau_f,\tau_i)|x_i\!=\!0\rangle_{\Wi}
&=&\int_{x(\tau_i)=x_i=0}^{x(\tau_f)=x_f=0}\![dx]\~e^{-\frac{1}{\hbar}S[\Bar{X}_{\tau_1}+x]}\cr
&\equi{\rm WKB}&{}_{\Wf}\langle x_f\!=\!0|
u_{\tau_1}(\tau_f,\tau_i)|x_i\!=\!0\rangle_{\Wi}
\~e^{-\frac{1}{\hbar}S[\Bar{X}_{\tau_1}]},
\label{1posoverlap01}
\eea
where
\bea
{}_{\Wf}\langle x_f\!=\!0|
u_{\tau_1}(\tau_f,\tau_i)|x_i\!=\!0\rangle_{\Wi}
&=&\int_{x(\tau_i)=x_i=0}^{x(\tau_f)=x_f=0}\![dx]
\exp\left\{-\frac{1}{2\hbar}
\int_{\tau_i}^{\tau_f}\! d\tau
\left(\dot{x}^2+x^2V^{\prime\prime}(\Bar{X}_{\tau_1})\right)\right\} \cr
&\stackrel{\e{1positionoverlap02}}{\propto}&
{\rm Det}\left\{-\partial^2_{\tau}
+V^{\prime\prime}(\Bar{X}_{\tau_1}(\tau)) \right\}^{-1/2}
\label{1posoverlap02}
\eea
is a functional determinant of quantum fluctuations.
(In the WKB expansion \e{1posoverlap01} of the action \e{action01},
there are no terms linear in the quantum fluctuation $x$ due to the
EL \eq{el01} and the Dirichlet boundary conditions.
The action terms quadratic in $x$ make up the functional determinant 
\e{1posoverlap02}.)
In Section~\ref{1inst}, we derive the fact that a 1-instanton contribution
is\footnote{We use the shorthand notation $\tau_{fi}:=\tau_f\!-\!\tau_i$,
$\tau_{f1}:=\tau_f\!-\!\tau_1$, $\tau_{1i}:=\tau_1\!-\! \tau_i$, and so forth.}
\bea
{}_{\Wf}\langle e_f|U_{\tau_1}(\tau_f,\tau_i)|e_i\rangle_{\Wi}
&\equi{\e{1posoverlap01}}&
{}_{\Wf}\langle e_f|u_{\tau_1}(\tau_f,\tau_i)|e_i\rangle_{\Wi}
\~e^{-\frac{1}{\hbar}S[\Bar{X}_{\tau_1}]}\cr
&=& Ke^{-\frac{1}{\hbar}e_f\tau_{f1}-\frac{1}{\hbar}e_i\tau_{1i}}
\label{1energyoverlap01}
\eea
before the integration over the collective coordinate $\tau_1$ of the
1-instanton. The $K$ in \eq{1energyoverlap01} is 
\bea
K&=& \left(\frac{\sqrt{\frac{\omega_f}{\omega_i}}
+\sqrt{\frac{\omega_i}{\omega_f}}}{2}\right)^{-1/2}
\sqrt{\frac{\Bar{S}}{2\pi\hbar}} e^{-\frac{1}{\hbar}\Bar{S}} K_0 \cr
&\equi{\e{kayzero01}}&
\left(\frac{\omega_f}{\pi\hbar}\right)^{1/4}
\left(\frac{\omega_i}{\pi\hbar}\right)^{1/4}
\sqrt{A_iA_f\Bar{S}} e^{-\frac{1}{\hbar}\Bar{S}}\cr
&\equi{\e{ca01}}&
\left(\frac{\omega_f^3}{\pi\hbar}\right)^{1/4}
\left(\frac{\omega_i^3}{\pi\hbar}\right)^{1/4}
\sqrt{C_iC_f} e^{-\frac{1}{\hbar}\Bar{S}}.
\label{kay01}
\eea
In \eq{kay01}, $K_0$ is a Gelfand-Yaglom (GY) quotient \e{kayzero00} of
functional determinants while the $C_{i/f}$ and $A_{i/f}$ denote asymptotically
defined amplitudes of a 1-instanton, see \eqs{asymp01}{0mode02} for precise
definitions. The 1-instanton formulas \es{1energyoverlap01}{kay01} are the
main results. 

\subsection{$N$-multi-instanton}
\label{ninstsec}

\noi
The dilute instanton gas approximation is valid when
\beq
\omega_i,\~ \omega_f\~\gg\~\tau_{fi}^{-1}
\qquad\text{and}\qquad\Bar{S}\~\gg\~\hbar. \label{diga01}
\eeq
The dilute $N$-instanton gas approximation with well-separated
collective coordinates $\tau_1\leq\ldots\leq\tau_N$ is given in terms of
1-instanton contributions:
\bea
{}_{\Wf}\langle e_f|U(\tau_f,\tau_i)|e_i\rangle_{\Wi}
&=&
\sum_N\int_{\tau_i\equiv\tau_{\frac{1}{2}}\leq\tau_1\leq\tau_{\frac{3}{2}}\leq
\ldots\leq\tau_{N-\frac{1}{2}}\leq\tau_N\leq\tau_{N+\frac{1}{2}}\equiv\tau_f}
\!d\tau_1\ldots d\tau_N\sum_{e_{\frac{3}{2}},\ldots,e_{N-\frac{1}{2}}} \cr
&&{}_{N+\frac{1}{2}}\langle e_{N+\frac{1}{2}}|
U_{\tau_N}(\tau_{N+\frac{1}{2}},\tau_{N-\frac{1}{2}})|
e_{N-\frac{1}{2}}\rangle_{N-\frac{1}{2}}
\~\ldots 
{}_{\frac{3}{2}}\langle e_{\frac{3}{2}}|
U_{\tau_1}(\tau_{\frac{3}{2}},\tau_{\frac{1}{2}})|
e_{\frac{1}{2}}\rangle_{\frac{1}{2}}
\~.\label{energyoverlap01}
\eea
The times $\tau_{\frac{1}{2}}\leq\ldots\leq\tau_{N+\frac{1}{2}}$ (with
half-integer subscripts) are a short-hand notation for $N\!+\!1$ finite time
intervals, each of which the system is approximately in a SHO ground state of
a well. They occur before, between or after the $N$ instanton transitions
$\tau_1\leq\ldots\leq\tau_N$ (with integer subscripts).

\section{Main application: same-level asymmetric double well}
\label{oddevensec}

\noi
In a double well, an $N$-multi-instanton can either start and end in the
same well (zig-zagging an even number $N\!=\!2n$ of times, see
Subsection~\ref{evensec}); or in different wells (zig-zagging an odd number
$N\!=\!2n+1$ of times, see next Subsection~\ref{oddsec}).

\noi
Let $\tau_1,\ldots,\tau_N\in[\tau_i,\tau_f]\equiv[\tau_0,\tau_{N+1}]$ denote
the collective coordinates of the individual instantons and anti-instantons.
Let $t_k:=\tau_k-\tau_{k-1}$ be the difference between the subsequent
collective coordinates.

\subsection{Odd instanton sector $N=2n+1$}
\label{oddsec}

\begin{figure}[ht]
\setlength{\unitlength}{.5ex} 
\centering 
\begin{picture}(180,60) 
\put(10,10){\vector(1,0){160}} 
\put(172,10){$\tau$}  
\put(10,10){\vector(0,1){40}} 
\put(8,52){$X$}
\put(2,20){$X_i$}
\put(9,20){\line(1,0){2}} 
\put(2,40){$X_f$}
\put(9,40){\line(1,0){2}} 
\put(20,6){$\tau_i$}
\put(20,9){\line(0,1){2}} 
\put(40,6){$\tau_1$}
\put(40,9){\line(0,1){2}} 
\put(60,6){$\tau_2$}
\put(60,9){\line(0,1){2}} 
\put(80,6){$\tau_3$}
\put(80,9){\line(0,1){2}} 
\put(120,6){$\tau_{N-1}$}
\put(120,9){\line(0,1){2}} 
\put(140,6){$\tau_N$}
\put(140,9){\line(0,1){2}} 
\put(160,6){$\tau_f$}
\put(160,9){\line(0,1){2}} 
\put(29,13){$t_1$}
\put(49,13){$t_2$}
\put(69,13){$t_3$}
\put(128,13){$t_N$}
\put(145,13){$t_{N+1}$}
\put(20,20){\line(1,0){20}} 
\put(40,20){\line(0,1){20}} 
\put(40,40){\line(1,0){20}} 
\put(60,20){\line(0,1){20}} 
\put(60,20){\line(1,0){20}} 
\put(96,20){$\ldots$}
\put(120,20){\line(1,0){20}} 
\put(140,20){\line(0,1){20}} 
\put(140,40){\line(1,0){20}} 
\end{picture}
\caption{A possible $N$-instanton path between an initial (``$i$") and
a final (``$f$") well in the odd instanton sector $N=2n+1$, where
$n\in\mathbb{N}_0$.} 
\label{figodd} 
\end{figure}
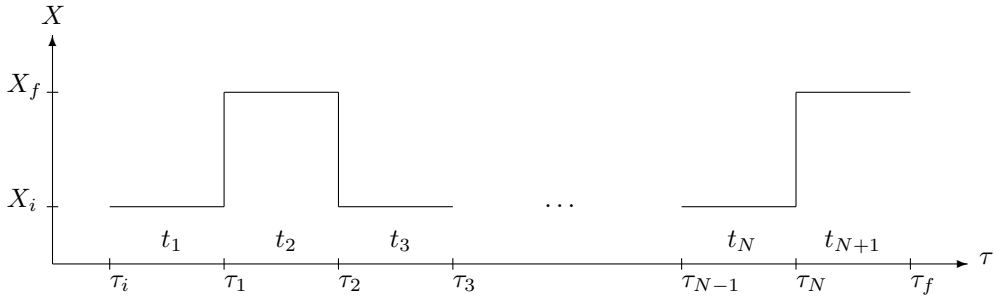

\noi
Let $T_i=\sum_{k\text{ odd}} t_k$ and $T_f=\sum_{k\text{ even}} t_k$
denote the total time spent in the initial (``$i$") and
the final (``$f$") well, respectively.
In the odd instanton sector, \cf Fig.~\ref{figodd},
the overlap is \cite{bmw94,dunne00,rivero02,vybornyi14,jlr20}
\bea
{}_{\Wf}\langle e_f|U(\tau_f,\tau_i)|e_i\rangle_{\Wi}
&\equi{\e{1energyoverlap01}+\e{energyoverlap01}}&
\sum_{n\in\mathbb{N}_0}K^N\int_{\tau_i\equiv\tau_0\leq\tau_1\leq
\ldots\leq\tau_N\leq\tau_{N+1}\equiv\tau_f}\!d\tau_1\ldots d\tau_N
\~e^{-\frac{1}{\hbar}e_fT_f-\frac{1}{\hbar}e_iT_i}\cr
&=&\sum_{n\in\mathbb{N}_0}K^N \int_{\mathbb{R}_+^{N+3}}\!dt_1\ldots dt_{N+1}dT_idT_f
\~e^{-\frac{1}{\hbar}e_fT_f-\frac{1}{\hbar}e_iT_i}\cr
&&\qquad\delta\left(\sum_{k\text{ odd}} t_k\-T_i\right)
\delta\left(\sum_{k\text{ even}} t_k\-T_f\right)\delta(T_i+T_f-\tau_{fi})\cr
&=&\sum_{n\in\mathbb{N}_0}K^{2n+1} \int_{\mathbb{R}_+^2}\!dT_idT_f\~
\frac{T_i^n}{n!}\frac{T_f^n}{n!}e^{-\frac{1}{\hbar}e_fT_f-\frac{1}{\hbar}e_iT_i}
\delta(T_i+T_f-\tau_{fi})\cr
&\equi{\e{laplacetransfodd01}}&
\frac{k}{\sqrt{1+4k^2}}\left\{e^{-\frac{1}{\hbar}E_-\tau_{fi}}
-e^{-\frac{1}{\hbar}E_+\tau_{fi}}\right\}. \label{eueodd}
\eea
The last line in \eq{eueodd} displays the minor miracle that
the quantum system is a 2-level system with energy levels\footnote{
The Taylor expansion is
$\sqrt{1+4k^2}-1
=\sum_{n\in \mathbb{N}} \begin{pmatrix} 1/2 \cr n\end{pmatrix} (2k)^{2n}
=2k^2-\sum_{n=2}^{\infty}\frac{(2n-3)!!}{n!} (-2k^2)^n.$ Note that the
division by $e_{fi}$ in \eq{energyodd01} is a removable singularity, not a
physical singularity for $e_{fi}\to 0$. Also note that in the dilute 
instanton gas approximation \e{diga01}, the instanton corrections are much 
smaller than the SHO zero-point energies \e{freq01}.} 
\bea
E_{\pm}&\equi{\e{detodd01}}&
\frac{e_f+e_i}{2}\pm \sqrt{\left(\frac{e_{fi}}{2}\right)^2+(\hbar K)^2}, \qquad 
e_{fi}\~:=\~e_f-e_i,\cr
&=& \frac{e_f+e_i}{2}\pm \frac{|e_{fi}|}{2}\sqrt{1+4k^2}, \qquad
k\~:=\~\frac{\hbar K}{|e_{fi}|},\cr
&=&\begin{pmatrix}\max(e_i,e_f)\cr \min(e_i,e_f)\end{pmatrix} \pm |\Delta E|,
\qquad  \Delta E\= \frac{e_{fi}}{2}\left(\sqrt{1+4k^2}-1\right),\cr
&\equi{e_f\geq e_i}&\begin{pmatrix}e_f\cr e_i\end{pmatrix} \pm \Delta E.
\label{energyodd01}
\eea
The energy levels \e{energyodd01} are roots in the characteristic
polynomial
\beq
\begin{vmatrix}
e_i-E & \pm\hbar K \cr \pm\hbar K & e_f-E
\end{vmatrix}\=0.\label{detodd01}
\eeq
To see the last equality in \eq{eueodd}, we perform a Laplace transform
$\tau_{fi}\leftrightarrow E$, which reveals via a partial fraction decomposition
(PFD) that this is a 2-level system \cite{bmw94}:
\bea
\int_{\mathbb{R}_+}\!\frac{d\tau_{fi}}{\hbar}e^{\frac{1}{\hbar}E\tau_{fi}}
{}_{\Wf}\langle e_f|U(\tau_f,\tau_i)|e_i\rangle_{\Wi}
&\equi{\e{eueodd}}&\frac{1}{\hbar}\sum_{n\in\mathbb{N}_0}K^{2n+1}
\int_{\mathbb{R}_+^2}\!dT_idT_f\~ \frac{T_i^n}{n!}\frac{T_f^n}{n!}
e^{\frac{1}{\hbar}(E-e_f)T_f+\frac{1}{\hbar}(E-e_i)T_i}\cr
&=&\sum_{n\in\mathbb{N}_0} \frac{(\hbar K)^{2n+1}}{\{(e_f-E)(e_i-E)\}^{n+1}} \cr
&\equi{\begin{matrix}\text{\scriptsize geom.}\cr
\text{\scriptsize series}\end{matrix}}&
\frac{\hbar K}{(e_f-E)(e_i-E)-(\hbar K)^2} \cr
&\equi{\e{detodd01}}& \frac{\hbar K}{(E_+-E)(E_--E)} \cr
&\equi{\text{PFD}}& \frac{\hbar K}{E_+-E_-}
\left\{\frac{1}{E_--E}-\frac{1}{E_+-E} \right\} \cr
&\equi{\e{energyodd01}}&
\frac{k}{\sqrt{1+4k^2}}\left\{\frac{1}{E_--E}-\frac{1}{E_+-E} \right\}.
\label{laplacetransfodd01}
\eea

\subsection{Even instanton sector $N=2n$}
\label{evensec}

\begin{figure}[ht]
\setlength{\unitlength}{.5ex} 
\centering 
\begin{picture}(200,60) 
\put(10,10){\vector(1,0){180}} 
\put(192,10){$\tau$}  
\put(10,10){\vector(0,1){40}} 
\put(8,52){$X$}
\put(2,20){$X_i$}
\put(9,20){\line(1,0){2}} 
\put(2,40){$X_m$}
\put(9,40){\line(1,0){2}} 
\put(20,6){$\tau_i$}
\put(20,9){\line(0,1){2}} 
\put(40,6){$\tau_1$}
\put(40,9){\line(0,1){2}} 
\put(60,6){$\tau_2$}
\put(60,9){\line(0,1){2}} 
\put(80,6){$\tau_3$}
\put(80,9){\line(0,1){2}} 
\put(120,6){$\tau_{N-2}$}
\put(120,9){\line(0,1){2}} 
\put(140,6){$\tau_{N-1}$}
\put(140,9){\line(0,1){2}} 
\put(160,6){$\tau_N$}
\put(160,9){\line(0,1){2}} 
\put(180,6){$\tau_f$}
\put(180,9){\line(0,1){2}} 
\put(29,13){$t_1$}
\put(49,13){$t_2$}
\put(69,13){$t_3$}
\put(125,13){$t_{N-1}$}
\put(148,13){$t_N$}
\put(165,13){$t_{N+1}$}
\put(20,20){\line(1,0){20}} 
\put(40,20){\line(0,1){20}} 
\put(40,40){\line(1,0){20}} 
\put(60,20){\line(0,1){20}} 
\put(60,20){\line(1,0){20}} 
\put(96,20){$\ldots$}
\put(120,20){\line(1,0){20}} 
\put(140,20){\line(0,1){20}} 
\put(140,40){\line(1,0){20}} 
\put(160,20){\line(0,1){20}} 
\put(160,20){\line(1,0){20}}
\end{picture}
\caption{A possible $N$-instanton path between an initial (``$i$") and
an intermediate/middle (``$m$") well in the even instanton sector $N=2n$,
where $n\in\mathbb{N}$.} 
\label{figeven} 
\end{figure}
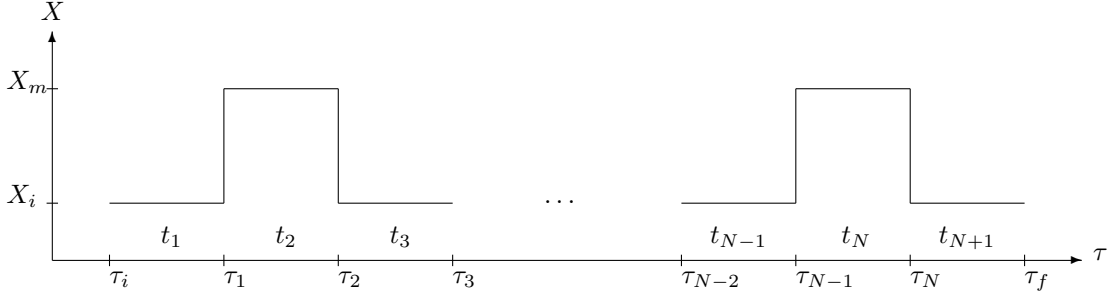

\noi
In the even instanton sector, \cf Fig.~\ref{figeven}, the initial and
final well are the same. The other well is called the intermediate/middle
(``$m$") well. Let $T_i=\sum_{k\text{ odd}} t_k$ and $T_m=\sum_{k\text{ even}} t_k$
denote the total time spent in the ``$i/m$'' well, respectively.
In the even instanton sector, the overlap is
\cite{bmw94,dunne00,rivero02,vybornyi14,jlr20}
\bea
{}_{\Wi}\langle e_i|U(\tau_f,\tau_i)|e_i\rangle_{\Wi}
&\equi{\e{1energyoverlap01}+\e{energyoverlap01}}&
e^{-\frac{1}{\hbar}e_i\tau_{fi}}
+\sum_{n\in\mathbb{N}}K^N\int_{\tau_i\equiv\tau_0\leq\tau_1\leq
\ldots\leq\tau_N\leq\tau_{N+1}\equiv\tau_f}\!d\tau_1\ldots d\tau_N
\~e^{-\frac{1}{\hbar}e_mT_m-\frac{1}{\hbar}e_iT_i}\cr
&=&e^{-\frac{1}{\hbar}e_i\tau_{fi}}
+\sum_{n\in\mathbb{N}}K^N \int_{\mathbb{R}_+^{N+3}}\!dt_1\ldots dt_{N+1}dT_idT_m
\~e^{-\frac{1}{\hbar}e_mT_m-\frac{1}{\hbar}e_iT_i}\cr
&&\qquad\qquad\qquad\delta\left(\sum_{k\text{ odd}} t_k\-T_i\right)
\delta\left(\sum_{k\text{ even}} t_k\-T_m\right)\delta(T_i+T_m-\tau_{fi})\cr
&=&e^{-\frac{1}{\hbar}e_i\tau_{fi}}
+\sum_{n\in\mathbb{N}}K^{2n}\int_{\mathbb{R}_+^2}\!dT_idT_m
\~\frac{T_i^n}{n!}\frac{T_m^{n-1}}{(n-1)!}
e^{-\frac{1}{\hbar}e_mT_m-\frac{1}{\hbar}e_iT_i}\delta(T_i+T_m-\tau_{fi})\cr
&\equi{e_m\geq e_i}&\frac{1+(1+4k^2)^{-1/2}}{2}e^{-\frac{1}{\hbar}E_-\tau_{fi}} 
\~+\~\frac{1-(1+4k^2)^{-1/2}}{2}e^{-\frac{1}{\hbar}E_+\tau_{fi}}. 
\label{eueeven}
\eea
The last line in \eq{eueeven} displays the minor miracle that
the quantum system is a 2-level system with energy levels 
\bea
E_{\pm}&\equi{\e{deteven01}}&
\frac{e_m+e_i}{2}\pm \sqrt{\left(\frac{e_{mi}}{2}\right)^2+(\hbar K)^2}, \qquad 
e_{mi}\~:=\~e_m-e_i,\cr
&=& \frac{e_m+e_i}{2}\pm \frac{|e_{mi}|}{2}\sqrt{1+4k^2}, \qquad
k\~:=\~\frac{\hbar K}{|e_{mi}|},\cr
&=&\begin{pmatrix}\max(e_i,e_m)\cr \min(e_i,e_m)\end{pmatrix} \pm |\Delta E|,
\qquad  \Delta E\= \frac{e_{mi}}{2}\left(\sqrt{1+4k^2}-1\right),\cr
&\equi{e_m\geq e_i}&\begin{pmatrix}e_m\cr e_i\end{pmatrix} \pm \Delta E.
\label{energyeven01}
\eea
The energy levels \e{energyeven01} are roots in the characteristic
polynomial
\beq
\begin{vmatrix}
e_i-E & \pm\hbar K \cr \pm\hbar K & e_m-E
\end{vmatrix}\=0.\label{deteven01}
\eeq
To see the last equality in \eq{eueeven}, we perform a Laplace transform
$\tau_{fi}\leftrightarrow E$, which reveals via a partial fraction decomposition
(PFD) that this is a 2-level system \cite{bmw94}:
\bea
\int_{\mathbb{R}_+}\!\frac{d\tau_{fi}}{\hbar}e^{\frac{1}{\hbar}E\tau_{fi}}
{}_{\Wi}\langle e_i|U(\tau_f,\tau_i)|e_i\rangle_{\Wi}
&\equi{\e{eueeven}}&\frac{1}{e_i-E}
+\frac{1}{\hbar}\sum_{n\in\mathbb{N}}K^{2n} \int_{\mathbb{R}_+^2}\!dT_idT_m
\~\frac{T_i^n}{n!}\frac{T_m^n}{(n-1)!}
e^{\frac{1}{\hbar}(E-e_m)T_m+\frac{1}{\hbar}(E-e_i)T_i}\cr
&=&\sum_{n\in\mathbb{N}_0} \frac{(\hbar K)^{2n}}{(e_i-E)^{n+1}(e_m-E)^n} \cr
&\equi{\begin{matrix}\text{\scriptsize geom.}\cr
\text{\scriptsize series}\end{matrix}}&
\frac{e_m-E}{(e_i-E)(e_m-E)-(\hbar K)^2}\cr
&\equi{\e{deteven01}}& \frac{e_m-E}{(E_+-E)(E_--E)}\cr
&\equi{\text{PFD}}&
\frac{1}{2}\left\{\frac{1}{E_--E}+\frac{1}{E_+-E} \right\}
\~+\~\frac{e_{mi}}{4(E_+-E_-)}\left\{\frac{1}{E_--E}-\frac{1}{E_+-E} \right\}\cr
&\equi{e_m\geq e_i}&\frac{1+(1+4k^2)^{-1/2}}{2}\frac{1}{E_--E} 
\~+\~\frac{1-(1+4k^2)^{-1/2}}{2}\frac{1}{E_+-E}.
\label{laplacetransfeven01}
\eea

\section{Functional determinant of 1-instanton}
\label{1inst}

\noi
In this Section~\ref{1inst} we calculate the functional determinant
\e{1posoverlap02} of a 1-instanton solution $X_{\tau_1}$.

\subsection{Infinite time period $\tau_{fi}=\infty$}

\noi
We will assume that the Hessian of the action is non-negative, and that the
pertinent functional determinant has a discrete non-negative spectrum. Up to
now we have assumed that the time period $\tau_{fi}:=\tau_f-\tau_i=\infty$ is
infinite, so that there is a time translation symmetry and a zero-mode
\cite{coleman85,cc77}
\bea
x_0(\tau)&:=&\dot{\Bar{X}}_{\tau_1}/\sqrt{\Bar{S}}\cr\cr
\e{sbar01}&\Downarrow&\cr\cr
\int_{\tau_i}^{\tau_f}\! d\tau~x_0(\tau)^2&=&1,\label{0mode01}
\eea
which is normalizable, \cf \eq{asymp01}.
The quantum trajectory has an expansion in terms of eigenfunctions
\bea
X&=&\Bar{X}_{\tau_1}+ \sum_{n\in\mathbb{N}_0} x_n c_n,\cr
&\Downarrow&\cr
dX&=&\dot{\Bar{X}}_{\tau_1}d\tau_1+ \sum_{n\in\mathbb{N}_0} x_n dc_n, \label{dex02}
\eea
where $\tau_1$ is the collective time coordinate of the 1-instanton.
In \eq{dex02}, $x_n(\tau)$ is the real orthonormal eigenfunction for the
eigenvalue $\lambda_n$, 
\beq
\int_{\tau_i}^{\tau_f}\! d\tau~x_n(\tau)x_m(\tau)\=\delta_{nm},
\qquad x_n(\tau_i)\=0\=x_n(\tau_f),\label{orthonormal03}
\eeq
and $c_n$ are the corresponding path integral integration variables with path
integral measure 
\beq
[dX]\=\prod_{n\in\mathbb{N}_0}\frac{dc_n}{\sqrt{2\pi\hbar}}.
\label{pathintmeasure01}
\eeq
Integrating over both the zero-mode $c_0$ and the collective coordinate
$\tau_1$ would result in overcounting, \cf \eqs{0mode01}{dex02}. Using the
normalization \e{pathintmeasure01} of the $c_0$-measure, we replace the
$c_0$-integration with the equivalently normalized $\tau_1$-integration
\beq
\int_{\mathbb{R}}\!\frac{dc_0}{\sqrt{2\pi\hbar}}\qquad\longrightarrow\qquad
\int_{\tau_i}^{\tau_f}\!d\tau_1 \sqrt{\frac{\Bar{S}}{2\pi\hbar}}.
\label{c0int01}
\eeq
The 1-instanton quantum overlap becomes 
\bea
\left(\frac{\omega_f}{\pi\hbar}\right)^{1/4}
\~{}_{\Wf}\langle e_f|u_{\tau_1}(\tau_f,\tau_i)|e_i\rangle_{\Wi}
\~\left(\frac{\omega_i}{\pi\hbar}\right)^{1/4}
&\stackrel{\e{gaussianwavefct01}}{\approx}&
{}_{\Wf}\langle x_f\!=\!0|u_{\tau_1}(\tau_f,\tau_i)|x_i\!=\!0\rangle_{\Wi}\cr
&\equi{\e{c0int01}}&
\sqrt{\frac{\Bar{S}}{2\pi\hbar}}
{\rm Det}^{\prime}\left\{-\partial^2_{\tau}
+V^{\prime\prime}(\Bar{X}_{\tau_1}(\tau)) \right\}^{-1/2}
\label{1positionoverlap02}
\eea
before the integration over the collective coordinate $\tau_1$. The first
prime symbol on the \rhs of \eq{1positionoverlap02} indicates that the
zero-eigenvalue has been removed from the functional determinant.

\subsection{Finite but large time period $\tau_{fi}<\infty$}

\noi
From now on we assume that $\omega_{f/i}^{-1}\ll\tau_{fi}<\infty$, so that the
time translation symmetry is broken, and that the lowest eigenvalue
$\lambda_0>0$ is strictly positive, although exponentially small. Define a
quotient quantity
\bea
K^{-2}_0&:=&\frac{{\rm Det}^{\prime}\left\{-\partial^2_{\tau}
+V^{\prime\prime}(\Bar{X}_{\tau_1}(\tau)) \right\}}{{\rm Det}\left\{
-\partial^2_{\tau} +V^{\prime\prime}_{(0)}(\Delta\tau_1) \right\}}
\~\equi{\e{diffop01}}\~
\frac{{\rm Det}^{\prime}\{\hat{A}\}}{{\rm Det}\{\hat{A}_{(0)}\}}\cr
&=&\frac{{\rm Det}\{\hat{A}\}}{\lambda_0{\rm Det}\{\hat{A}_{(0)}\}}
\~\equi{\text{GY thm}}\~\frac{\psi_0(\tau_f)}{\lambda_0\psi^{(0)}_0(\tau_f)}.
\label{kayzero00}
\eea
In \eq{kayzero00}, the functional determinant is defined relative to a
reference model (notationally adorned with a ``$(0)$").
In Section~\ref{relfuncdetsec}, we calculate the relative functional
determinant using the Gelfand-Yaglom (GY) theorem by examining the zero-modes
of the corresponding differential operators
\beq \hat{A}\~:=\~-\partial^2_{\tau} +V^{\prime\prime}(\Bar{X}_{\tau_1}(\tau))
\qquad\text{and}\qquad
\hat{A}_{(0)}\~:=\~-\partial^2_{\tau}  +V^{\prime\prime}_{(0)}(\Delta\tau_1).
\label{diffop01}
\eeq

\subsection{Step function reference model}

\noi
As a reference model, we choose a time-dependent harmonic oscillator:
\beq
V_{(0)}(X)\~:=\~\frac{1}{2} \sum_{\pm}\theta(\pm\Delta\tau_1)
\omega_{\pm} (X-X_{\pm})^2, \qquad 
\Delta\tau_1\~:=\~\tau-\tau_1.\label{refpot01}
\eeq
A system with $m$ wells typically has asymptotically a higher-order power-law
potential $V(X)\sim X^{2m}$ for $X\to \pm\infty$. One may wonder whether the GY
theorem works with an asymptotically quadratic reference potential?
Fortunately, in practice, this is granted by the pertinent Fredholm and
Sturm-Liouville theories.

\noi
The classical solution (satisfying Dirichlet boundary conditions)
is a step function \wrt time:
\beq
\Bar{X}^{(0)}_{\tau_1}(\tau)
\=\sum_{\pm}\theta(\pm\Delta\tau_1)X_{\pm}
\~:=\~\theta(\tau\-\tau_1)X_f\~+\~\theta(\tau_1\-\tau)X_i,
\qquad \tau\~\neq\~\tau_1.\label{clasrefsol01}
\eeq
Note that the velocity
$\dot{\Bar{X}}^{(0)}_{\tau_1}(\tau)=(X_f\-X_i)\delta(\Delta\tau_1)$ is a Dirac
delta distribution, which renders the classical action infinite/mathematically
ill-defined: $S_{(0)}[\Bar{X}^{(0)}_{\tau_1}]=\infty$. Luckily, the classical
solution $\Bar{X}^{(0)}_{\tau_1}$ does not actually enter the calculation of
the main 1-instanton formulas \es{1energyoverlap01}{kay01}. Only the
quantum fluctuations will be important.

\noi
The Hessian is also a step function \wrt time:
\beq
V^{\prime\prime}_{(0)}(\Delta\tau_1)
\~\equi{\e{refpot01}}\~\sum_{\pm}\theta(\pm\Delta\tau_1)\omega_{\pm}
\~:=\~\theta(\tau\-\tau_1)\omega_f\~+\~\theta(\tau_1\-\tau)\omega_i.
\label{hessian01}\eeq
Fortunately, the Hessian \e{hessian01} does not depend on $X$, nor the
classical solution \e{clasrefsol01}.
The functional determinant becomes
\bea
{\rm Det}\{\hat{A}_{(0)}\}^{-1/2}
&\equi{\e{diffop01}}&
{\rm Det}\left\{-\partial^2_{\tau}
+V^{\prime\prime}_{(0)}(\Delta\tau_1) \right\}^{-1/2}\cr
&=&\int_{x(\tau_i)=x_i=0}^{x(\tau_f)=x_f=0}\![dx]
\exp\left\{-\frac{1}{2\hbar}
\int_{\tau_i}^{\tau_f}\! d\tau
\left(\dot{x}(\tau)^2+x(\tau)^2V^{\prime\prime}_{(0)}(\Delta\tau_1)\right)\right\}
\cr
&=&{}_{\Wf}\langle x_f\!=\!0|u^{(0)}_{\tau_1}(\tau_f,\tau_i)|
x_i\!=\!0\rangle_{\Wi}\cr
&\stackrel{\e{gaussianwavefct01}}{\approx}&
\left(\frac{\omega_f}{\pi\hbar}\right)^{1/4}
\~{}_{\Wf}\langle e_f|u^{(0)}_{\tau_1}(\tau_f,\tau_i)|e_i\rangle_{\Wi}
\~\left(\frac{\omega_i}{\pi\hbar}\right)^{1/4},
\label{detdot02}
\eea
where
\bea
\~{}_{\Wf}\langle e_f|u^{(0)}_{\tau_1}(\tau_f,\tau_i)|e_i\rangle_{\Wi}
&\equi{\begin{matrix}\text{\scriptsize compl.}
\cr\text{\scriptsize rela.}\end{matrix}}&
\int_{\mathbb{R}}\! dx_1  \~ {}_{\Wf}\langle e_f|
u^{(0)}(\tau_f,\tau_1)|x_1\rangle_{\Wf}
\~{}_{\Wi}\langle x_1|u^{(0)}(\tau_1,\tau_i)|e_i\rangle_{\Wi}\cr
&\equi{\rm SHO}&
\int_{\mathbb{R}}\! dx_1  \~ e^{-\frac{1}{\hbar}e_f\tau_{f1}}
\~{}_{\Wf}\langle e_f|x_1\rangle_{\Wf}
\~ {}_{\Wi}\langle x_1|e_i\rangle_{\Wi}e^{-\frac{1}{\hbar}e_i\tau_{1i}}\cr
&\equi{\e{gaussianwavefct01}}&
e^{-\frac{1}{\hbar}e_f\tau_{f1}-\frac{1}{\hbar}e_i\tau_{1i}}
\left(\frac{\omega_f}{\pi\hbar}\right)^{1/4}
\left(\frac{\omega_i}{\pi\hbar}\right)^{1/4}
\int_{\mathbb{R}}\! dx_1  e^{-\frac{1}{2\hbar}(\omega_i+\omega_f)x_1^2}\cr
&\equi{\begin{matrix}\text{\scriptsize Gauss.}
\cr\text{\scriptsize int.}\end{matrix}}&
\left(\frac{\omega_f}{\pi\hbar}\right)^{1/4}
\left(\frac{\omega_i}{\pi\hbar}\right)^{1/4}
\sqrt{\frac{2\pi\hbar}{\omega_i+\omega_f}}
e^{-\frac{1}{\hbar}e_f\tau_{f1}-\frac{1}{\hbar}e_i\tau_{1i}}\cr
&=&\left(\frac{\sqrt{\frac{\omega_f}{\omega_i}}
+\sqrt{\frac{\omega_i}{\omega_f}}}{2}\right)^{-1/2}
e^{-\frac{1}{\hbar}e_f\tau_{f1}-\frac{1}{\hbar}e_i\tau_{1i}}.
\label{1energyoverlapdot02}
\eea
Combining eqs.\ \e{1positionoverlap02}, \e{kayzero00},
\es{detdot02}{1energyoverlapdot02} yield the main 1-instanton formulas
\es{1energyoverlap01}{kay01}. $\Box$

\section{Relative functional determinant via Gelfand-Yaglom (GY) theorem}
\label{relfuncdetsec}

\noi
In this Section~\ref{relfuncdetsec} we calculate formula \e{kayzero01} for
the $K_0$-quotient \e{kayzero00} via the GY theorem by considering zero-modes
\es{psinot02}{psinotnot01} for the corresponding differential operators
\e{diffop01}.

\subsection{Zero-mode for the $\hat{A}$ operator}
\label{zeromode}

\noi
In the infinite time limit $\tau_{fi}=\infty$ the 1-instanton velocity profile
is a normalizable zero-mode:
\bea
x_0(\tau)&\stackrel{\e{0mode01}}{:=}&\dot{\Bar{X}}/\sqrt{\Bar{S}}\cr
&\stackrel{\e{asymp01}}{\approx}&
A_{\pm}e^{-\omega_{\pm}|\Delta\tau_1|}
\quad\for\quad
\omega_{\pm}|\Delta\tau_1|\~\gg\~1.\label{0mode02}
\eea
The asymptotic amplitudes $C_{\pm}$ and $A_{\pm}$ from
\eqs{asymp01}{0mode02} satisfy
\beq
|A_{\pm}|\= C_{\pm}\omega_{\pm}/\sqrt{\Bar{S}}.\label{ca01}
\eeq
Because we assume that the zero-mode is the lowest eigenvalue, it is nodeless
according to the node theorem. Therefore, we may assume that the amplitudes
$A_{\pm}\in \mathbb{R}$ are real with the same sign. In light of the
normalization \e{0mode01}, we expect the amplitudes
$A_{\pm}\sim{\cal O}(\omega_{\pm}^{1/2})$. The 2nd-order differential operator
\e{diffop01} has a linearly independent 2nd (non-normalizable) zero-mode
$y_0(\tau)$. The Wronskian $W$ is a non-zero constant, which we select to be of
order $W\sim{\cal O}(\omega_{\pm}^2)$ for practical purposes:
\bea
0\~\neq\~W&:=&W(x_0,y_0)\~:=\~x_0\dot{y}_0-y_0\dot{x}_0\label{wronskian01}\cr
&\stackrel{\e{0mode02}}{\approx}&
A_{\pm}e^{-\omega_{\pm}|\Delta\tau_1|}(\dot{y}_0\pm\omega_{\pm}y_0)\cr
&=&A_{\pm}e^{-2\omega_{\pm}|\Delta\tau_1|}\frac{d}{d\tau}(e^{\omega_{\pm}|\Delta\tau_1|}y_0)
\quad\for\quad
\omega_{\pm}|\Delta\tau_1|\~\gg\~1.\cr
&\Downarrow&\cr
\exists B_{\pm}:\qquad y_0(\tau)
&\approx&\pm \frac{W}{2A_{\pm}\omega_{\pm}}e^{\omega_{\pm}|\Delta\tau_1|}
+B_{\pm}e^{-\omega_{\pm}|\Delta\tau_1|}
\quad\for\quad
\omega_{\pm}|\Delta\tau_1|\~\gg\~1.\label{whynot01}
\eea
For dimensional reasons, we expect 
\beq
\int_{\tau_i}^{\tau_f}\! d\tau~y_0(\tau)^2
\~\sim\~
{\cal O}\left(e^{2\omega_{\pm}|\tau_{\pm}-\tau_1|}\right)
\qquad\text{and}\qquad
\int_{\tau_i}^{\tau_f}\! d\tau~x_0(\tau)y_0(\tau)
\~\sim\~
{\cal O}\left(\omega_{\pm}|\tau_{\pm}-\tau_1|\right).
\eeq
By shifting $y_0\to y_0-\frac{B_-}{A_-}x_0$ we may assume wlog.\ that $B_{-}=0$.
The general zero-mode is a linear combination
\bea
\psi_0(\tau)&=& a x_0(\tau) + b y_0(\tau)\label{psinot01}\cr
&\equi{\e{ic01}}&\frac{1}{2A_i\omega_i}e^{\omega_i\tau_{1i}}x_0(\tau) 
+\frac{A_i}{W}e^{-\omega_i\tau_{1i}}y_0(\tau),
\label{psinot02}\eea
where $a,b\in\mathbb{C}$ are two constants.
In the 2nd equality of \eq{psinot02}, we used the GY initial conditions (IC):
\bea
\begin{pmatrix} 0 \cr 1\end{pmatrix}
\=\begin{pmatrix} \psi_0(\tau_i) \cr \dot{\psi}_0(\tau_i)\end{pmatrix} 
&\equi{\e{0mode02}+\e{whynot01}}& \begin{pmatrix}   aA_ie^{-\omega_i\tau_{1i}}
-b\frac{W}{2A_i\omega_i}e^{\omega_i\tau_{1i}} \cr
aA_i\omega_ie^{-\omega_i\tau_{1i}}
+b\frac{W}{2A_i}e^{\omega_i\tau_{1i}} \end{pmatrix} \cr
&=& \begin{pmatrix} 1 & \omega_i^{-1} \cr \omega_i & -1 \end{pmatrix}
\begin{pmatrix} aA_ie^{-\omega_i\tau_{1i}} \cr
-\frac{bW}{2A_i}e^{\omega_i\tau_{1i}}\end{pmatrix}\cr
&\Downarrow&\cr
\begin{pmatrix} aA_ie^{-\omega_i\tau_{1i}} \cr
-\frac{bW}{2A_i}e^{\omega_i\tau_{1i}}\end{pmatrix}
&=&\frac{1}{2}\begin{pmatrix} \omega_i^{-1} \cr -1\end{pmatrix},\qquad
\tau_{1i}\~:=\~\tau_1-\tau_i.\label{ic01}
\eea
The retarded Green function is
\bea
G(\tau,\tau^{\prime}) &=& \frac{\theta(\tau-\tau^{\prime})}{W}
\left\{ x_0(\tau)y_0(\tau^{\prime})- y_0(\tau)x_0(\tau^{\prime})\right\}\cr
&\Downarrow&\cr
\partial_{\tau}G(\tau,\tau^{\prime})
&=& \frac{\theta(\tau-\tau^{\prime})}{W}\left\{ \dot{x}_0(\tau)y_0(\tau^{\prime})
-\dot{y}_0(\tau)x_0(\tau^{\prime})\right\}\cr
&\Downarrow&\cr
\hat{A}G(\tau,\tau^{\prime})
&\equi{\e{diffop01}}&
\left\{-\partial^2_{\tau} +V^{\prime\prime}(\Bar{X}_{\tau_1}(\tau)) \right\}
G(\tau,\tau^{\prime})
\=\delta(\tau\-\tau^{\prime}).
\eea
The retarded Green function $G$ can be used to solve the inhomogeneous
initial value problems (IVP)
\beq
\left. \begin{matrix}
\hat{A}\psi(\tau)&=&j(\tau)\cr
\psi(\tau_i)&=&\psi_i\cr
\dot{\psi}(\tau_i)&=&\dot{\psi}_i
\end{matrix}\right\}
\qquad\Rightarrow \qquad
\psi(\tau)\=\psi_i\~+\~(\tau-\tau_i)\dot{\psi}_i\~+\~
\int_{\tau_i}^{\tau_f}\!d\tau^{\prime}~G(\tau,\tau^{\prime})j(\tau^{\prime}).
\eeq
The eigenfunction $\psi_{\lambda}$ satisfies
\bea
\left\{\hat{A} -\lambda\right\}\psi_{\lambda}(\tau)&=&0
\quad\wedge\quad
\psi_{\lambda}(\tau_i)\=0
\quad\wedge\quad
\dot{\psi}_{\lambda}(\tau_i)\=1
\label{eigenvalueeq01}\cr
&\Downarrow&\cr
{\cal O}(\lambda^0):\qquad\qquad
\hat{A}\psi_0(\tau)&=& 0,\label{eigenvalueeq02}\cr
{\cal O}(\lambda^1):\qquad
\hat{A}\left.\frac{\partial\psi_{\lambda}(\tau)}{\partial\lambda}
\right|_{\lambda=0}&=& \psi_0(\tau).\label{eigenvalueeq03}
\eea
The lowest eigenvalue $\lambda_0>0$ is assumed to be exponentially small,
and it satisfies the final condition (FC) 
\bea
\psi_{\lambda_0}(\tau_f)&=&0\cr
&\Downarrow&\cr
-\frac{\psi_0(\tau_f)}{\lambda_0}
&=&\frac{\psi_{\lambda_0}(\tau_f)-\psi_0(\tau_f)}{\lambda_0}
\~\approx\~
\left.\frac{\partial\psi_{\lambda}(\tau_f)}{\partial\lambda}\right|_{\lambda=0}\cr
&\equi{\e{eigenvalueeq03}}&
\underbrace{\left.\frac{\partial\psi_{\lambda}(\tau_i)}{\partial\lambda}
\right|_{\lambda=0}}_{=0}
\~+\~(\tau_f-\tau_i)
\underbrace{\left.\frac{\partial\dot{\psi}_{\lambda}(\tau_i)}{\partial\lambda}
\right|_{\lambda=0}}_{=0}
\~+\~\int_{\tau_i}^{\tau_f}\!d\tau~G(\tau_f,\tau)\psi_0(\tau)\cr
&\equi{\e{psinot02}}&\frac{1}{W}\int_{\tau_i}^{\tau_f}\!d\tau
\left\{ A_fe^{-\omega_f\tau_{f1}}y_0(\tau)
-\left(\frac{W}{2A_f\omega_f}e^{\omega_f\tau_{f1}}+
B_fe^{-\omega_f\tau_{f1}}\right)x_0(\tau)\right\}\cr
&&\qquad
\left\{\frac{1}{2A_i\omega_i}e^{\omega_i\tau_{1i}}x_0(\tau) 
+\frac{A_i}{W}e^{-\omega_i\tau_{1i}}y_0(\tau)\right\}\cr
&\stackrel{\e{whynot01}}{\approx}&
-\frac{1}{4A_iA_f\omega_i\omega_f}e^{\omega_i\tau_{1i}+\omega_f\tau_{f1}}.
\label{quotient01}
\eea

\subsection{Reference model: zero-mode for the $\hat{A}_{(0)}$ operator}
\label{zeromodenot}

\noi
The zero-eigenvalue equation is
\beq
\hat{A}_{(0)}\psi^{(0)}_0(\tau)\~\equi{\e{diffop01}}\~
\left\{-\partial^2_{\tau} +V^{\prime\prime}_{(0)}(\Delta\tau_1) \right\}
\psi^{(0)}_0(\tau)\=0.
\label{0modeeqnot01}\eeq
The general continuous solution (with a continuous derivative) of the
homogeneous 2nd-order linear ODE \e{0modeeqnot01} is:
\beq
\psi^{(0)}_0(\tau)\equi{\e{hessian01}+\e{0modeeqnot01}}
\sum_{\pm}\theta(\pm\Delta\tau_1)\left\{A\cosh(\omega_{\pm}\Delta\tau_1)
+ B\omega_{\pm}^{-1}\sinh(\omega_{\pm}\Delta\tau_1)\right\},
\label{psinotnot01}
\eeq
where $A,B\in\mathbb{C}$ are two constants.
 The GY initial conditions (IC) are:
\bea
\begin{pmatrix} 0 \cr 1\end{pmatrix}
\=\begin{pmatrix} \psi^{(0)}_0(\tau_i) \cr
\dot{\psi}^{(0)}_0(\tau_i)\end{pmatrix} 
&\equi{\e{psinotnot01}}& \begin{pmatrix} A\cosh(\omega_i\tau_{1i})
-B\omega_i^{-1}\sinh(\omega_i\tau_{1i}) \cr  
-A\omega_i\sinh(\omega_i\tau_{1i})
+B\cosh(\omega_i\tau_{1i}) \end{pmatrix}\cr
&=& \begin{pmatrix} \cosh(\omega_i\tau_{1i}) &
-\omega_i^{-1}\sinh(\omega_i\tau_{1i}) \cr
-\omega_i\sinh(\omega_i\tau_{1i}) & \cosh(\omega_i\tau_{1i}) \end{pmatrix}
\begin{pmatrix} A \cr B\end{pmatrix}\cr
&\Downarrow&\cr
\begin{pmatrix} A \cr B\end{pmatrix}
&=&\begin{pmatrix} \omega_i^{-1}\sinh(\omega_i\tau_{1i}) \cr
\cosh(\omega_i\tau_{1i}) \end{pmatrix}, \qquad
\tau_{1i}\~:=\~\tau_1-\tau_i.\label{icnot01}
\eea
So at the final time 
\bea
\psi^{(0)}_0(\tau_f)&\equi{\e{psinotnot01}+\e{icnot01}}&
\omega_i^{-1}\sinh(\omega_i\tau_{1i})\cosh(\omega_f\tau_{f1}) \cr
&&+\omega_f^{-1}\sinh(\omega_f\tau_{f1})\cosh(\omega_i\tau_{1i}) \cr 
&\approx&\frac{\omega_i^{-1}+\omega_f^{-1}}{4}
e^{\omega_i\tau_{1i}+\omega_f\tau_{f1}}.\label{psinotnotf01}
\eea

\subsection{Quotient $K_0$} 

\noi
The GY theorem now yields the quotient
\bea
K^2_0&\stackrel{\e{kayzero00}}{:=}&
\frac{\lambda_0{\rm Det}\{\hat{A}_{(0)}\}}{{\rm Det}\{\hat{A}\}}
\~\equi{\text{GY thm}}\~\frac{\lambda_0\psi^{(0)}_0(\tau_f)}{\psi_0(\tau_f)}\cr
&\equi{\e{quotient01}+\e{psinotnotf01}}&
\frac{\omega_i^{-1}+\omega_f^{-1}}{4} 4A_iA_f\omega_i\omega_f
\=A_iA_f(\omega_i+\omega_f).
\label{kayzero01}
\eea
$\Box$

\section{Examples} 
\label{exsec}

\subsection{Symmetric double well}

\noi
Let us first see that our formulas reproduce the well-known symmetric double
well \cite{coleman85,vzns82,rajaraman87,altlandsimons10,rattazzi11,paranjape22}

\bea
V(X)&=&\frac{\lambda}{4!}(X^2-a^2)^2\cr
&\Downarrow&\cr
V^{\prime}(X)&=&\frac{\lambda}{3!}(X^2-a^2)X\cr
&\Downarrow&\cr
V^{\prime\prime}(X)&=&\frac{\lambda}{2}\left(X^2-\frac{a^2}{3}\right)\cr
&\Downarrow&\cr
\omega^2&=&V^{\prime\prime}(\pm a)
\=\frac{\lambda a^2}{3}.\label{2pot01}
\eea
The assumption that the SHO ground state energy is much smaller than the
barrier potential yields the following condition:
\bea
\frac{\hbar \omega}{2}\~\ll\~ V(0)
&\equi{\e{2pot01}}&\frac{\lambda a^4}{4!}
\~\equi{\e{2pot01}}\~\frac{3\omega^4}{8\lambda}\cr
&\Downarrow&\cr
\hbar\lambda &\ll& \omega^3.\label{2pert01}
\eea
The 1-instanton solution from the ``$-$" well to the ``$+$" well satisfies:
\beq
\Bar{S}\=\int_{[-a,a]} \! dX~\sqrt{2V(X)}
\=\frac{2\omega a^2}{3}
\~\equi{\e{2pot01}}\~
\underbrace{\frac{2\omega^3}{\lambda}}_{\text{non-pert.}}
\~\stackrel{\e{2pert01}}{\gg}\~\hbar.
\eeq
where the instanton solutions are
\bea
\Bar{X}&\equi{\e{energycons00}}&
\pm a\tanh\left(\frac{\omega}{2}\Delta\tau_1\right)\cr
&\Downarrow&\cr
\dot{\Bar{X}}&=&
\pm \frac{a\omega}{2}\cosh^{-2}\left(\frac{\omega}{2}\Delta\tau_1\right)\cr
&\Downarrow&\cr
x_0&\equi{\e{0mode01}}&\dot{\Bar{X}}/\sqrt{\Bar{S}}
\=\pm \frac{1}{2}\sqrt{\frac{3\omega}{2}}
\cosh^{-2}\left(\frac{\omega}{2}\Delta\tau_1\right)\cr
&\Downarrow&\cr
A_{\pm}&\equi{\e{0mode02}}&\sqrt{6\omega}\cr
&\Downarrow&\cr
K_0&\equi{\e{kayzero01}}&2\sqrt{3}\omega\cr
&\Downarrow&\cr
K&\equi{\e{kay01}}&\sqrt{\frac{\Bar{S}}{2\pi\hbar}}
e^{-\frac{1}{\hbar}\Bar{S}} K_0.
\eea
The energies of the 1st excited state $\psi_1$ and the ground state $\psi_0$
are
\beq
 E_{1/0}\equi{\e{energyodd01}+\e{energyeven01}}
 \frac{\hbar\omega}{2}\pm \hbar K,
\eeq
respectively. The overlaps are
\bea
{}_{+}\langle e_+|U(\tau_f,\tau_i)|e_-\rangle_{-}&\equi{\e{eueodd}}& 
\frac{1}{2}\left\{e^{-\frac{1}{\hbar}E_0\tau_{fi}}
-e^{-\frac{1}{\hbar}E_1\tau_{fi}}\right\} ,\cr
{}_{+}\langle e_+|U(\tau_f,\tau_i)|e_+\rangle_{+}
&\equi{\e{eueeven}}& 
\frac{1}{2}\left\{e^{-\frac{1}{\hbar}E_0\tau_{fi}}
+e^{-\frac{1}{\hbar}E_1\tau_{fi}}\right\} .\label{2overlap01}
\eea
On the other hand, the overlaps can be expanded in energy eigenstates 
\beq
\langle X\!=\!a|U(\tau_f,\tau_i)|X\!=\!\pm a\rangle
\=\sum_{n\in\mathbb{N}_0}\langle X\!=\!a|E_n\rangle e^{-\frac{1}{\hbar}E_n\tau_{fi}}
\langle E_n|X\!=\!\pm a\rangle.\label{2overlap02}
\eeq
A comparison of eqs.\ \e{gaussianwavefct01}, \es{2overlap01}{2overlap02}
leads to the following values for the wavefunctions,
\beq
\begin{array}{lcccccl}
\text{ground state}&\psi_0(\pm a)&=&\langle \pm a\mid E_0\rangle 
&\approx&\frac{1}{\sqrt{2}}\left(\frac{\omega}{\pi\hbar}\right)^{1/4}
&\text{symmetric/0 nodes},\cr
\text{1st excited state}&\psi_1(\pm a)&= &\langle \pm a\mid E_1\rangle 
&\approx&\pm\frac{1}{\sqrt{2}}\left(\frac{\omega}{\pi\hbar}\right)^{1/4}
&\text{antisymmetric/1 node},
\end{array}
\eeq
modulo a phase convention.

\subsection{Same-level triple well}

\noi
A simple example of a same-level triple well at $X\in\{0,\pm a\}$ is the
potential \cite{rajchel91,lkyplpy96,al04,dsu20,pomajbo25}
\bea
V(X)&=&\frac{\lambda}{2}X^2(X^2-a^2)^2\cr
&\Downarrow&\cr
V^{\prime}(X)&=&\lambda(3X^4-4a^2X^2+a^4)X\cr
&\Downarrow&\cr
V^{\prime\prime}(X)&=&\lambda(15X^4-12a^2X^2+a^4)\cr
&\Downarrow&\cr
\text{Middle well:}\qquad\qquad  \omega_m^2&=&V^{\prime\prime}(0)
\=\lambda a^4\cr
\text{Left/right well:}\qquad  \omega_{\ell/r}^2&=&V^{\prime\prime}(\pm a)
\=4\lambda a^4\=(2\omega_m)^2.\label{3pot01}
\eea
Note that the SHO frequency $\omega_{\ell/r}=2\omega_m$ of the outer wells is
twice that of the middle well, so a subsystem [of the triple well] is an
example of a same-level asymmetric double well.

\noi
We will focus on the 3 lowest energy states; ignoring the higher states.
The ground state $\psi_0$ is a nodeless symmetric wavefunction located mainly
in the middle well, while the 1st (2nd) excited state is an antisymmetric
(symmetric) wavefunction $\psi_1$ ($\psi_2$) mainly located in the outer wells
with 1 (2) node(s), respectively. Owing to the parity symmetry, the 1st
excited state energy
\beq E_1\=\frac{\hbar\omega_r}{2}\=\hbar\omega_m\eeq 
does not receive instanton corrections. However, the ground state and the 2nd
excited state do mix.

\noi
The assumption that the SHO ground state energy is much smaller than the
barrier potential yields the following condition:
\bea
\frac{\hbar \omega_m}{2}\~\ll\~ \frac{\lambda a^6}{2}&\equi{\e{3pot01}}&
\frac{\omega_m^3}{2\sqrt{\lambda}}\cr
&\Downarrow&\cr
\hbar\sqrt{\lambda} &\ll& \omega_m^2.\label{3pert01}
\eea
The 1-instanton solution from the middle (``$m$") well to the right (``$r$")
well satisfies:
\bea
\Bar{S}&=&\int_{[0,a]} \! dX~\sqrt{2V(X)}
\=\sqrt{\lambda}\int_{[0,a]} \! dX~X(a^2-X^2)\cr
&=&\frac{\sqrt{\lambda}a^4}{4}
\~\equi{\e{3pot01}}\~
\frac{\omega_ma^2}{4}
\~\equi{\e{3pot01}}\~
\underbrace{\frac{\omega_m^2}{4\sqrt{\lambda}}}_{\text{non-pert.}}
\~\stackrel{\e{3pert01}}{\gg}\~\hbar.
\eea
where the instanton solution is
\bea
\Bar{X}&\equi{\e{energycons00}}&
a\left(1 +e^{-2\omega_m\Delta\tau_1}\right)^{-1/2}\cr
&\Downarrow&\cr
\dot{\Bar{X}}&=& \omega_ma\left(e^{\frac{2}{3}\omega_r\Delta\tau_1}
+e^{-\frac{2}{3}\omega_m\Delta\tau_1}\right)^{-3/2}\cr
&\Downarrow&\cr
x_0&\equi{\e{0mode01}}&\dot{\Bar{X}}/\sqrt{\Bar{S}}
\=2\sqrt{\omega_m}\left(e^{\frac{2}{3}\omega_r\Delta\tau_1}
+e^{-\frac{2}{3}\omega_m\Delta\tau_1}\right)^{-3/2}\cr
&\Downarrow&\cr
A_{\ell/m/r}&\equi{\e{0mode02}}&2\sqrt{\omega_m}\cr
&\Downarrow&\cr
K_0&\equi{\e{kayzero01}}&2\sqrt{3}\omega_m\cr
&\Downarrow&\cr
K&\equi{\e{kay01}}&\left(\frac{2\sqrt{2}}{3} \right)^{1/2}
\sqrt{\frac{\Bar{S}}{2\pi\hbar}} e^{-\frac{1}{\hbar}\Bar{S}} K_0.
\eea
The energies of the 2nd excited state $\psi_2$ and the ground state $\psi_0$
are
\bea
 E_{2/0}&\equi{\e{energyodd01}+\e{energyeven01}}&
 \frac{\hbar}{2}\begin{pmatrix}\omega_r\cr
 \omega_m\end{pmatrix} \pm \Delta E, \cr  
\Delta E&\equi{\e{energyodd01}+\e{energyeven01}}&
\frac{\hbar\omega_m}{2}\left(\sqrt{1+8k^2}-1\right),\cr
k&\equi{\e{energyodd01}+\e{energyeven01}}&\frac{2K}{\omega_m}.
\eea
The overlaps are
\bea
{}_{r}\langle e_r|U(\tau_f,\tau_i)|e_m\rangle_{m}&\equi{\e{eueodd}}& 
\frac{k}{\sqrt{1+8k^2}}\left\{e^{-\frac{1}{\hbar}E_0\tau_{fi}}
-e^{-\frac{1}{\hbar}E_2\tau_{fi}}\right\} , \cr
{}_{m}\langle e_m|U(\tau_f,\tau_i)|e_m\rangle_{m}&\equi{\e{eueeven}}& 
\frac{1+(1+8k^2)^{-1/2}}{2}e^{-\frac{1}{\hbar}E_0\tau_{fi}} 
\~+\~\frac{1-(1+8k^2)^{-1/2}}{2}e^{-\frac{1}{\hbar}E_2\tau_{fi}} , \cr
{}_{r}\langle e_r|U(\tau_f,\tau_i)|e_{r/\ell}\rangle_{r/\ell}&\equi{\e{eueeven}}& 
\pm\frac{1}{2}e^{-\frac{1}{\hbar}E_1\tau_{fi}} 
\~+\~\frac{1-(1+8k^2)^{-1/2}}{4}e^{-\frac{1}{\hbar}E_0\tau_{fi}} \cr
&& +\frac{1+(1+8k^2)^{-1/2}}{4}e^{-\frac{1}{\hbar}E_2\tau_{fi}} .
\label{3overlap01}
\eea
On the other hand, the overlaps can be expanded in energy eigenstates 
\bea
\langle X\!=\!-a|U(\tau_f,\tau_i)|X\!=\! 0\rangle
&=&\sum_{n\in\mathbb{N}_0}\langle X\!=\!-a|E_n\rangle e^{-\frac{1}{\hbar}E_n\tau_{fi}}
\langle E_n|X\!=\! 0\rangle,\cr
\langle X\!=\!0|U(\tau_f,\tau_i)|X\!=\! 0\rangle
&=&\sum_{n\in\mathbb{N}_0}\langle X\!=\!0|E_n\rangle e^{-\frac{1}{\hbar}E_n\tau_{fi}}
\langle E_n|X\!=\! 0\rangle,\cr
\langle X\!=\!-a|U(\tau_f,\tau_i)|X\!=\!\pm a\rangle
&=&\sum_{n\in\mathbb{N}_0}\langle X\!=\!-a|E_n\rangle e^{-\frac{1}{\hbar}E_n\tau_{fi}}
\langle E_n|X\!=\!\pm a\rangle.\label{3overlap02}
\eea
A comparison of eqs.\ \e{gaussianwavefct01}, \es{3overlap01}{3overlap02}
leads to the following values for the wavefunctions,
\beq
\begin{array}{lccccl}
\text{ground state}:&\psi_0(0)&=&\langle 0\mid E_0\rangle 
&\approx&\sqrt{\frac{1+(1+8k^2)^{-1/2}}{2}}
\left(\frac{\omega_m}{\pi\hbar}\right)^{1/4},\cr
&\psi_0(\pm a)&=&\langle \pm a\mid E_0\rangle 
&\approx&\frac{\sqrt{1-(1+8k^2)^{-1/2}}}{2}
\left(\frac{\omega_r}{\pi\hbar}\right)^{1/4},\cr
\text{2nd excited state}:&\psi_2(\pm a)
&=&\langle \pm a\mid E_2\rangle 
&\approx&\frac{\sqrt{1+(1+8k^2)^{-1/2}}}{2}
\left(\frac{\omega_r}{\pi\hbar}\right)^{1/4},\cr
&\psi_2(0)&=&\langle 0\mid E_2\rangle 
&\approx&-\sqrt{\frac{1-(1+8k^2)^{-1/2}}{2}}
\left(\frac{\omega_m}{\pi\hbar}\right)^{1/4},\cr
\text{1st excited state}:&\psi_1(\pm a)
&=&\langle \pm a\mid E_1\rangle 
&\approx&\pm\frac{1}{\sqrt{2}}
\left(\frac{\omega_r}{\pi\hbar}\right)^{1/4},\cr
&\psi_1(0)&=&\langle 0\mid E_1\rangle &=&0,
\end{array}
\eeq
modulo phase conventions.

\noi
Note that the $K^2$-factor (and $k^2$-factor) have effectively been
multiplied by a combinatorical factor $2$ because the middle (``$m$") well
has 2 possible instanton-bounces. Here, an instanton-bounce is a 2-instanton,
also known as a bion \cite{dsu20}.

\section*{Acknowledgement}

The author thanks Filip Pomajbo for fruitful discussions. The work of
K.B.\ was supported by the Czech Science Foundation (GACR) under grant no.\
GA26-22343S.

\end{document}